\documentclass{article}
\usepackage{ijcai24}
\usepackage{times}
\usepackage{soul}
\usepackage{url}
\usepackage[hidelinks]{hyperref}
\usepackage[utf8]{inputenc}
\usepackage[small]{caption}
\usepackage{graphicx}
\usepackage{amsmath}
\usepackage{amsthm}
\usepackage{booktabs}
\usepackage{algorithm}
\usepackage{algorithmic} 
\usepackage[T1]{fontenc}
\usepackage{subfig}
\usepackage{makeidx}

\title{EEG Based Generative Depression Discriminator}

\author{
Ziming Mao$^1$
\and
Hao Wu$^1$\thanks{Corresponding author}\and
Yongxi Tan$^1$\and
Yuhe Jin$^1$\\
\affiliations
$^1$School of Computer Science and Technology, Beijing Institute of Technology\\
\emails
mzimo@foxmail.com,
wuhao123@bit.edu.cn,
\{849373525,741619413\}@qq.com
}

\begin{document}
\maketitle
\begin{abstract}
Depression is a very common but serious mood disorder.In this paper,
We built a generative detection network(GDN) in accordance with three physiological laws. Our aim is that we expect the neural network to learn the relevant brain activity based on the EEG signal and, at the same time, to regenerate the target electrode signal based on the brain activity.
We trained two generators, the first one learns the characteristics of depressed brain activity, and the second one learns the characteristics of control group's brain activity. In the test, a segment of EEG signal was put into the two generators separately, if the relationship between the EEG signal and brain activity conforms to the characteristics of a certain category, then the signal generated by the generator of the corresponding category is more consistent with the original signal.
Thus it is possible to determine the category corresponding to a certain segment of EEG signal. We obtained an accuracy of 92.30\% on the MODMA dataset and 86.73\% on the HUSM dataset. Moreover, this model is able to output explainable information, which can be used to help the user to discover possible misjudgments of the network.Our code will be released.
\end{abstract}

\section{Introduction}
In the present era, an great number of people are
suffering from depression. As revealed by WHO surveys,
an estimated 3.8\% of the population experience depression,
 including 5\% of adults (4\% among men and 6\% among women), and 
 5.7\% of adults older than 60 years.\cite{WHO}
Furthermore, the repercussions of depression are multifaceted,
posing significant risks to individuals,
it can co-occur with other serious medical illnesses, 
such as diabetes, cancer, heart disease, and Parkinson’s disease,
especially in midlife or older age\cite{NIMH}
However, the underlying mechanisms of the onset of depression
remain unclear\cite{Shen2021An,EEG_biomarker}
 ,and prevailing clinical identification of depression
  relies primarily on self-reported questionnaires and interviews.
  Meanwhile, the diagnostic outcomes provided by questionnaire 
 would come with the inherent drawback of subjectivity, and the
  diagnosis from psychiatrists exhibit variability based
   on their clinical specialization. \cite{BACHMANN2017391,Shen2021An}
Thus, there is a demand for diagnosing depression objectively.

There are a number of studies aimed at making depression judgements by discovering biomarkers(physiological characteristics that are both detectable
 and capable of distinguishing individuals with depression
  from normal controls\cite{EEG_biomarker})
  in the electroencephalogram (EEG),
because EEG can reveal complex brain functions such as cognition, emotion, attention, and memory\cite{EEG_and_Cognitive_Neuroscience}.
For example, \cite{Lee} discovered that the left brain alpha-band power was found to be higher in depressed patients.
\cite{Grin} discovered that the alpha, beta and theta bands at occipital and parietal areas were more active in depressed patients.
However, until now, we can't give out an established EEG biomarker to figure
 out whether a person is experiencing depression\cite{EEG_biomarker2},
what's more, it is expected that a single biomarker would not fully characterize it\cite{EEG_biomarker,EEG_biomarker2}

In recent years, the development of deep learning
 has been evident to all.
 Thus, many individuals are attempting to use EEG signals to train deep learning
  models aiming to achieve the goal of depression identification.
   For example,\cite{10182389} uses a GCN-based model to make classification, and
   \cite{10289973} uses 1D-CNN + MLP to perform depression recognition.

These deep learning networks all have one feature, 
their starting point is that the brain activity of depressed patients is different from normal controls,
and these differences can be reflected in the EEG of the two, 
then they use the deep learning network to learn the differences between the EEG signals of the depressed patients and those of the controls, 
so that the neural network can perform depression recognition.
They trained their network by using label '0' and '1'.
One of the biggest problems with doing this is that the EEG is a superposition of many brain activity discharges at the scalp\cite{EEG_Signal_Processing_and_Feature_Extraction}, 
and we can't guarantee that the differences between the EEG signals are entirely from the abnormal brain activity caused by the depression, if so, it is unlikely that we can't give out en established EEG biomarker\cite{EEG_biomarker2}.
Thus our experiments can only be based on the fact that the EEG signals contain information about the brain activity that is different. 
Therefore, we built a model based on this point, which was designed with the goal of not classifying EEG signals by their differences, 
as in previous studies, but reconstructing brain activity based on the EEG signals, 
and discriminating on the basis of their brain activity.
    
Before presenting our model,
we introduce three theoretical background for the design of the whole model.
\begin{figure}
    \centering
    \includegraphics[width=1\linewidth]{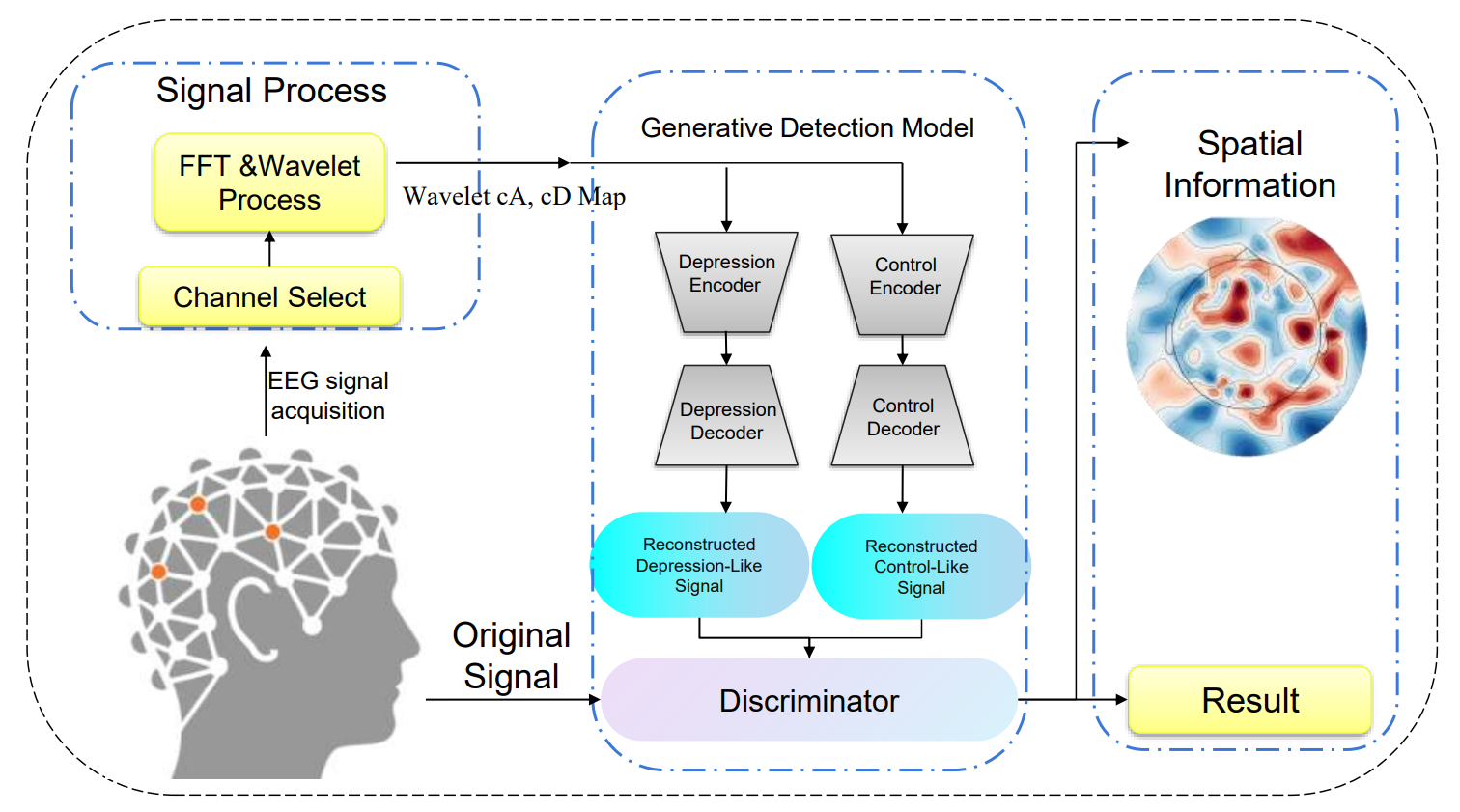}
    \caption{Generative Detection Network(GDN)}
    \label{1}
    \end{figure} 
\begin{itemize}
    \item EEG signals are not independent between different electrodes: 
      EEG signals are the superposition of brain activity discharges at the scalp,
       and the signals collected by each electrode do not only come from the brain
        area corresponding to that electrode, but are the integration
         of the discharges of many brain activities. \cite{EEG_Signal_Processing_and_Feature_Extraction}

    \item EEG signals contain information on brain function collaboration:
    EEG topography can somewhat represent changes
    in the overall writing pattern of functional brain activity
    This means that the EEG contains, to some extent,
    information about the brain activity. 
    \cite{EEG_Signal_Processing_and_Feature_Extraction}

    \item Depressed patients have differences in brain function from normal controls,
    and these differences can be reflected by EEG:
    There is a number of studies that address EEG biomarkers,
    and while these studies do not give an established biomarker on EEG,
    what they do have in common is that they suggest that there are differences
    between the EEG of depressed patients and that of normal controls,
    and they point out the differences they found. Also, according to\cite{EEG_biomarker2},
    the right-side hemisphere, frontal and parietal-occipital cortex are
    useful to detect depression using EEG signal.
    In conclusion, we take it as a prerequisite for our work that depressed patients have differences 
    in brain function from normal controls, 
    and these differences can be find in EEG.
\end{itemize}

    Based on these three physiological premises, we built a generative depression detection network(GDN fig\ref{1}) based on EEG.
    Not like those image generation models of DDPM\cite{DDPM}, GAN\cite{GAN} and VAE\cite{VAE}, which generate the final target from the noise, 
    our generative model generates the subject's EEG signals from the brain activity.
    The inputs to our network are
    several electrode signals\(D_{1}, D_{2}, \dots,D_{n}\)that are 'similar'
    to a target electrode signal\(D_{m}\),
    Then we use an Encoder to transform the input signals into brain activity which relate to the target electrode.
    Finally, we use a Decoder to reconstruct target electrode signal from brain activity,
    and the output is a piece of generated EEG data\(D^{'}\).
    According to the premise, we can learn that for a certain electrode, 
    its signal\(D_{m}\) comes from certain brain activity,
    and we expect that the deep learning network 
    can reconstruct this certain brain activity based on EEG signals similar to it,
    so that the generated EEG signals\(D_{m}^{'}\)
    will be able to close to the original signal\(D_{m}\). 
    If the difference between the reconstructed EEG data \(D_{m}^{'}\)
    and the original signal \(D_{m}\) is small, it means that the neural network
    has learnt the correct correlation between EEG signals and brain activity.

    As stated in our premise, there is a difference between the brain function
    of depressed patients and that of normal controls, so we train two generators, 
    the first generator with all the data from the depressed patients, 
    aiming to learn the characteristics of the brain function of the depressed patients,
    and similarly, the second generator with all the data from the controls, 
    aiming to learn the characteristics of the brain function of the normal controls. 
    When we need to make a test, we get a piece of EEG, 
    and putting this piece of EEG into each of the two generators, 
    then we are able to get two pieces of EEG signals. 
    If the difference between the EEG generated by the depression generator and
    the original EEG is smaller than the control's, which means the corresponding brain activity of this piece of signal is closer to that of a depressed person's brain activity, then we consider the subject to be depressed.
    If contrary, we consider it to be a normal control.

    Our model achieved high accuracy on both publicly available datasets(92.30\% accuracy on MODMA dataset\cite{MODMA} and 86.73\% accuracy on a HUSM dataset\cite{HUSM} and we succeeded in correctly judging each subject's category.
    Besides, our model can give some explainable information which is helpful in spotting the misjudgement of the model.
    
\section{Relate Works}

EEG signals can be categorized into five frequency bands:\(\alpha,\beta,\theta,\gamma,\delta.\)
These five frequency bands can reflect different physiological conditions.\(\alpha\)(8Hz-12Hz) reflects the inactivity and relaxation of the brain. The asymmetry of Alpha is related to the approach-withdraw model of brain activity.\(\beta\)(12Hz-27Hz) is associated with expectancy, anxiety, and an inwardly focused state of mind, and occurs mainly in the parietal and frontal regions.\(\theta\)(3Hz-8Hz) is associated with emotional processing in the brain, and it has a high differentiation of activity at the central electrode.\(\gamma\)(27Hz and above) is associated with attention and sensory systems.\(\delta\)(0.2Hz-3Hz), on the other hand, is associated with deep sleep and occurs mainly in the sleep state.\cite{EEG_and_Cognitive_Neuroscience,EEG_Signal_Processing_and_Feature_Extraction,EEG_yixue_review}

Many studies using EEG for depression discrimination have identified some of the more prominent physiologic features.\cite{Hosseinifard,Grin} found that
Alpha and other bands in depressive EEG are distinguished from normal EEG, with an increase in Alpha waves being the main feature of the depressed.
\cite{Lee} found an increase in Alpha waves and a decrease in Beta waves in the left hemisphere of depressed patients by quantitative EEG studies.
\cite{Acharya} concluded that EEG signals from the right half of the brain are more suitable for deep learning categorization than the left half of the brain.
In our experiments, we first filtered the signals using the discrete Fourier transform, selecting information in the 4-14 Hz band. This band corresponds to the Theta wave and Alpha wave of EEG, which is not only related to the emotional processing of the brain, but also to the emotional processing of the brain.

In recent years, there are many depression recognition methods utilizing CNN (Convolutional Neural Network) to process EEG signals, such as EEGNet, TS-SEFFT, and so on.
The EEGNet design a single CNN architecture to accurately classify EEG signals.\cite{EEGNet}
The TS-SEFFT model\cite{9495768} extracts temporal and spectral features through T-Conv units and spectral convolution units.
The MV-SGDC-RAFFNet\cite{Beihang} used a GCN based method to learn temporal, spectral, and
time–frequency features from EEG signals jointly.

Also, there are some studies using Machine Learning Method to make depression judgement.
\cite{Shen2021An} used a method based on mordify kernel-target alignment(mKTA), aiming to
select several channels to make judgement.
The major difference between our work and the literature is that we don't use a
certain model to find out the differences between the depressed and the control,
our target is to learning the hidden brain activity signature of depression and control,
and by generating EEG from brain activity, we can make a classification.

And our model is generative discriminative model, which derives its idea from VAE\cite{VAE}.
VAE is a generative model which can generate a picture from noise.
During the training process, it contains two parts, Encoder and Decoder. Encoder transforms the image input into a hidden vector. This vector invisibly contains the information that the picture possesses. The Decoder, on the other hand, transforms the hidden vector back into an image.
These processes are possible only if the network is able to learn the features of the input image and can reproduce them with the Decoder.
And the same is true for our model, although our model is not trying to generate EEG from noise. However, we want the model Encoder to learn the connection between the input signal and the hidden brain activity, and by means of the Decoder outputs the target signal based on the activity reconstructed by Encoder.By comparing the generated signal with the original signal, we are able to determine whether the network has correctly grasped the  characteristics of the brain activity corresponding to this piece of EEG signal.

\section{Generative Detection Network}
    \subsection[3.1]{net overview}

    The overall network architecture possesses four parts, the first part is the process of the signal input.
    The second part is a Encoder to perform the feature extraction.
    The third part is a Decoder to generate the EEG signals of the target electrodes based on the extracted features, and the fourth part is to output explainable information in the space domains based on the network information.

    In the first part, take the MODMA dataset as an example, for each piece of signal\(D \in \mathbf{R}^{(128, 2500)}\), we calculate the 'similar signals' for each electrode [see \ref{3.2}].We choose the 10 most similar electrodes,
    Thus, the signal\(D\) input to the network becomes $(128, 11, 2500)$, where $D[:, 0, :]$ is the original data of the corresponding electrode, and $D[:, 1:, :]$ is the information of 10 electrodes that are similar to the current electrode signal.

    When signal selection is performed, we first use FFT to extract the information of the input signal in the frequency band of 4~14Hz, and then process it using wavelet transform to obtain the \(cA,cD\) feeture of the signal, and take these feature as the  input of the network.
    
    In the second part, we noticed that the \(cA\) features and \(cD\) features of each signal represent completely different information, so we trained a network on the \(cA\) data and a network on the \(cD\) data.

    In the third part, we first weighted and summed the hidden vector extracted by the second layer network.After that, we similarly trained two fully-connected layers, which generate the \(cA\) and \(cD\) of the target reconstructed EEG signals respectively

    During the training process, we use the MSELoss to calculate the loss between the original signal and the generated signal.
    Whereas, during validation and testing, we first use inverse wavelet transform to
    get the generated signal, then compute the loss between the generated and the original.
    
\subsection[3.2]{Data Process}
\label{3.2}
        \subsubsection[3.2.1]{Similarity Calculation}
        
        There are various ways to calculate the similarity between signals, such as calculating the correlation coefficient, calculating the cosine similarity, calculating the convolution between signals, the
        Calculating Manhattan distance and Euclidean distance between signals, etc.\cite{channel}

        In this experiment, we used cosine similarity to calculate the similarity between different electrodes.
        Each segment of input EEG data\(D \in \mathbf{R}^{(128, 2500)}\) has 128 electrodes.
        For each electrode signal\(D_{a} \in D, a \in \{1, 2, \dots, 128\}\)
        We calculate the similarity of this electrode signal to the other signals according to the following equation:
        \[
            \rho_{ab} = {\sum_{i=1}^{n}{(D_{a}(i) \times D_{b}(i))}} \over {\sqrt{\sum_{i=1}^{n}{D_{a}(i)^2}}} \times \sqrt{\sum_{i=1}^{n}{D_{b}(i)^2}}\tag{1}    
        \]
        Where \(D_{b} \in D - \{D_{a}\}\),\(n = 2500\) is the total number of sampling points        

        After calculating the similarity between different electrodes of similarity, for each electrode, we select \(k\) electrodes with the largest cosine similarity,
        and compose the signal\(D_{a}\) into \(S_{a} \in \mathbf{R}^{(1, k + 1, 2500)}\), where \(S_{a}[1, 0, :]\) is the original signal of this electrode.
        Regarding the choice of \(k\), we did four sets of comparison tests, in which the values of \(k\) were taken as 5, 10, 15, and 20.
        We found that the network works best when k is taken as 10. The specific data can be viewed in the experimental section (plus a link)

        \subsubsection[3.2.3]{Introduction to Discrete Fourier Transform}
        DFT is an important means of transforming signals from time domain to frequency domain in signal processing.
        According to related research, \(\theta\) wave are related to emotions, and many previous studies have found that the alpha waves of depressed patients are different from those of normal patients.
        Therefore, we use discrete Fourier transform for filtering and select the information in the 4-14Hz band.
        During the experiment, we first use Fast Fourier Transform on the signal to transform it to the frequency domain with the following equation.
        \[
            X_{a}[k] = \sum_{t=0}^{n-1}T_{s}(t)D_{a}(t)\exp^{-j2\pi\frac{tk}{n}} 
        \]
        \[(k \in {0, 1, \dots, n-1}) \tag{2}\]
        where \(\forall D_{a} \in D\), \(T_{s}\) are hamming window functions
        \[T_{s} = 0.54 + (0.46 \times \cos(t)) \tag{3} \]
        \(X[k]\) is the amplitude of the component with frequency \(F = F_{s}\frac{k}{n}\).
        \(F_{s}\) is the sampling frequency.

        Afterwards, the amplitudes in the frequency band of 4-14 Hz are selected for inverse discrete Fourier transform, \(F_{s}\) is the sampling frequency, and
        \[
            D_{a}(t) = \frac{1}{T_{s}(t)n}\sum_{k=0}^{n-1}X_{a}[k]\exp^{j2\pi\frac{kt}{n}} \tag{4}
        \]
        This completes the filtering.

        After processing the signal using the Fast Fourier Transform, we selected information in the 4-14 Hz band, which corresponds in the EEG to
        $\alpha$ and $\theta$ bands, which, according to current research in academia, are associated with emotion and cognition.\cite{EEG_and_Cognitive_Neuroscience}
        
\subsubsection[3.2.4]{'db6' wavelet transform}
The wavelet transform is able to decompose the signal into components of different scales with different frequencies, thus enabling the capture of local features.
"db6" is a member of the Daubechies family of wavelet functions in the wavelet transform. We choose 'db6' wavelet basis function to process our signal\(D\), because ‘db6’ function performs well with non-stationary signals\cite{db6}.
\[
    \phi(a, \tau) = \frac{1}{\sqrt{t}}\sum_{0}^{t-1} D(t) \times \psi_{a, \tau}(t) \tag{5}
\]
where \(\psi\) is the wavelet basis filter. After wavelet transform processing, we can get the wavelet coefficients, \(cA\) and \(cD\), which represent the low-frequency and high-frequency portions of the signal.

\subsection[3.3]{Depression and Control Signal Generator}
        \subsubsection[3.3.1]{GDN Encoder}

        When the number of relevant electrodes is chosen to be 10, and the 'db6' wavelet family is selected, for a segment of the EEG signal\(D\) in the MODMA dataset.
        It is processed to become:
        \(D_{processed} \in \mathbf{R}^{(128, 2, 11, 1255)}\)
        The implication is that there are 128 electrodes for each segment of EEG data.
        For each channel, we have two associated signal wavelet coefficient maps \(S_{cA} \in \mathbf{R}^{(10, 1255)} \quad
        S_{cD} \in \mathbf{R}^{(10, 1255)}\)
        and two original electrode information sequences \(O_{cA} \in \mathbf{R}^{(1, 1255)} \quad O_{cD} \in \mathbf{R}^{(10, 1255)}\)
        Since the wavelet parameters cA and cD represent different meanings, we build two feature extraction networks\(Encoder_{cA},Encoder_{cD}\) for processing the cA parameter and cD parameter respectively.

        Each Encoder consists of 6 layers of feature nodes, the structure of each feature node is shown in Fig. For the input of each node
        For each node's input \(x\), a Batch Normalization is performed first, followed by a conv layer with kernel size=1, and a residual join is performed after the completion.
        After that, a residual join is performed. After that, it enters another conv layer and passes the convolution output backward as the input of the node in the next layer.
        The formula for the whole process is shown below.
        \[
            x = Conv(Conv(BN(x)) + BN(x)) \tag{6}
        \]

        The inputs of these two feature extraction networks are the wavelet coefficient maps of the correlated signals for each channel \(S \in \mathbf{R}^{(10, 1255)} \)
        And the output of the network is the extracted features\(S^{'} \in \mathbf{R}^{(1, 300)} \) which represent the hidden brain activity.
        After that we pass the two hidden vectors\(S_{cA}^{'},S_{cD}^{'}\) into the Decoder.
        \begin{figure}
            \centering
            \subfloat[Encoder Node]{
            \label{2}
            \includegraphics[scale=0.2]{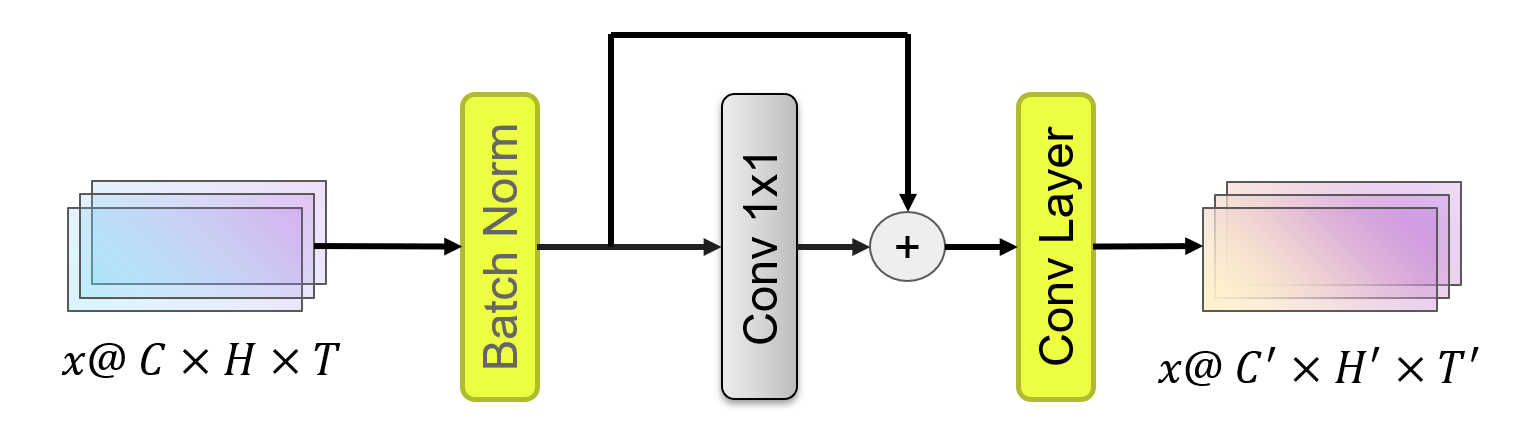}}
            \\
            \subfloat[Decoder Node]{
            \label{3}
            \includegraphics[scale=0.2]{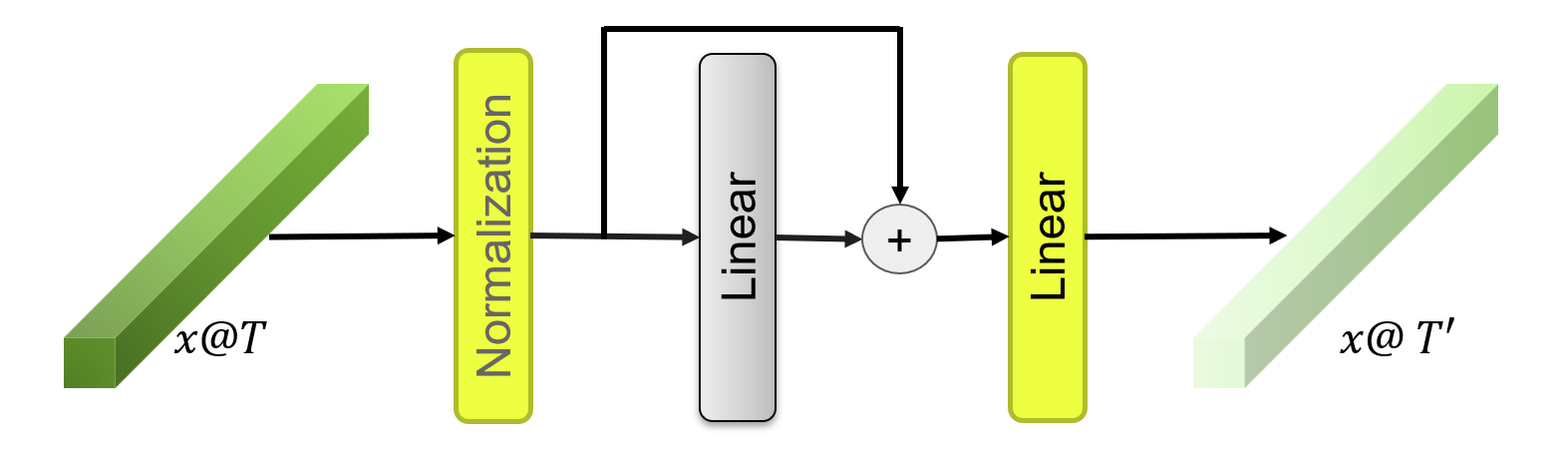} 
            }
            \end{figure}
        \subsubsection[3.4.2]{GDN Decoder}
            \begin{figure*}
    \centering
    \subfloat{
    \includegraphics[scale=0.23]{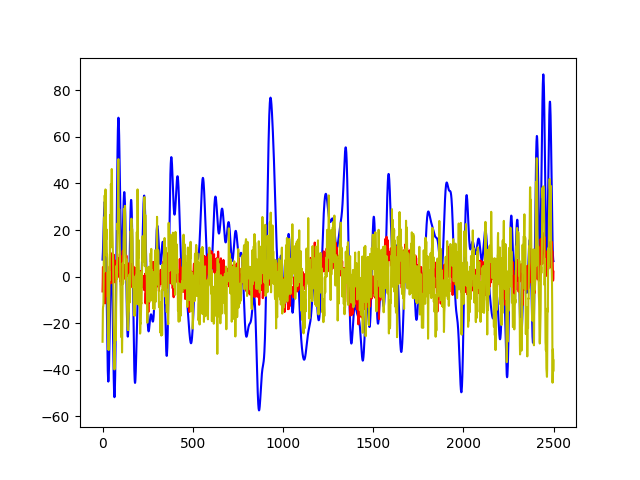} 
    }
    \subfloat{
    \includegraphics[scale=0.23]{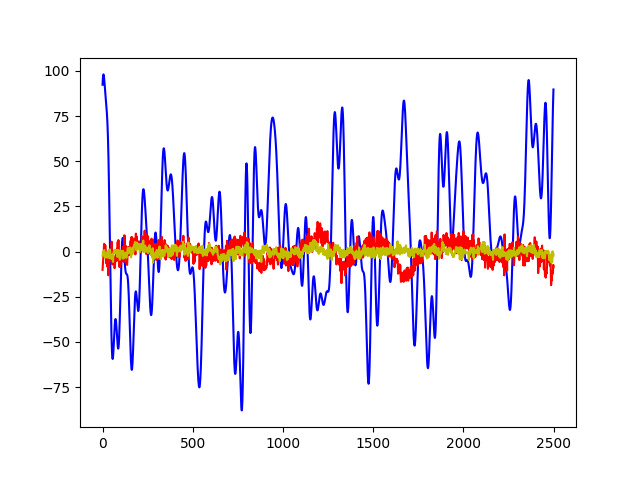} 
    }
    \subfloat{
    \includegraphics[scale=0.23]{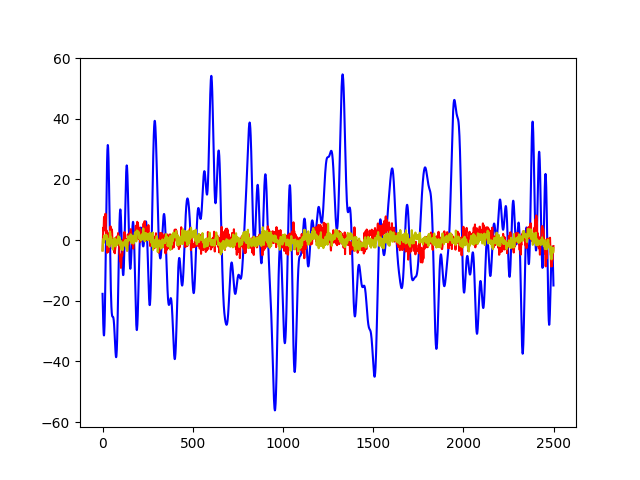}
    }
    \subfloat{
    \includegraphics[scale=0.23]{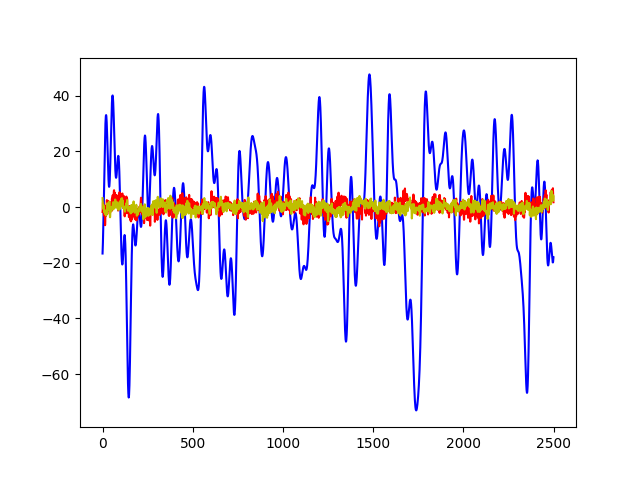}
    }
    \\
    \centering
    \subfloat{
    \includegraphics[scale=0.23]{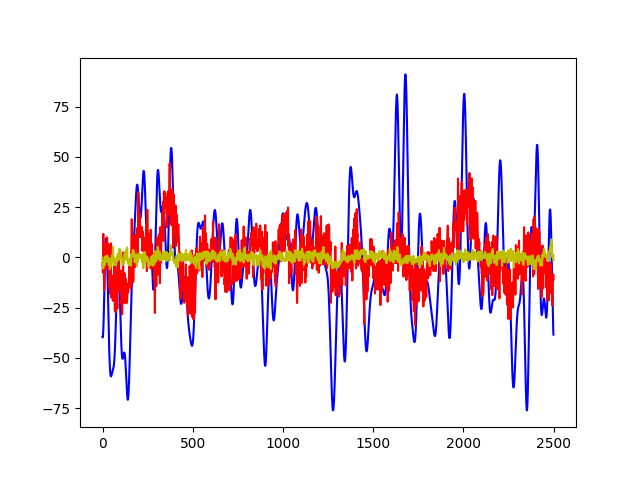} 
    }
    \subfloat{
    \includegraphics[scale=0.23]{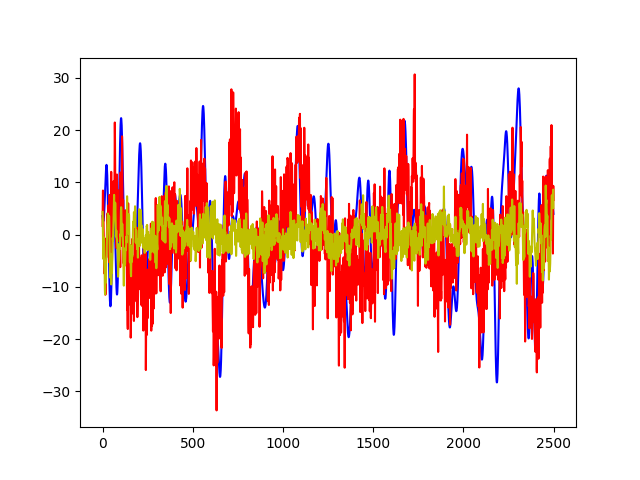} 
    }
    \subfloat{
    \includegraphics[scale=0.23]{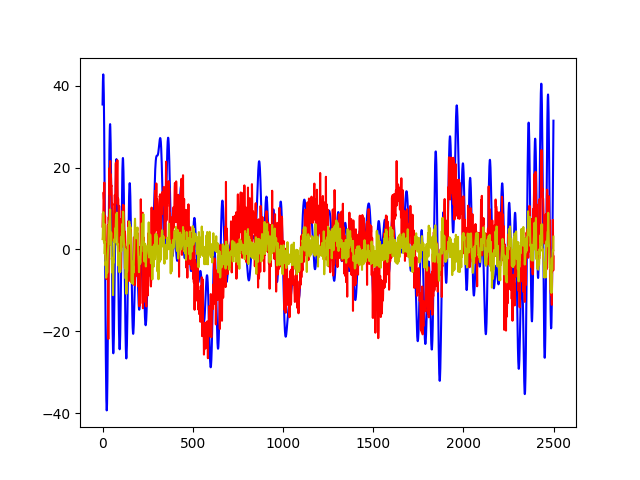}
    }
    \subfloat{
    \includegraphics[scale=0.23]{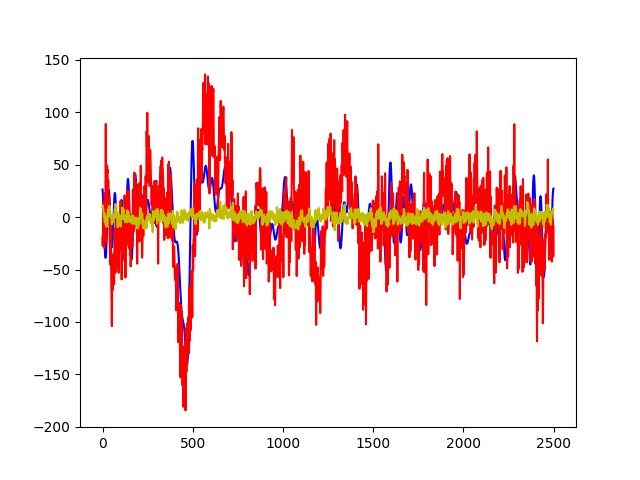}
    }
    \caption{These are comparisons of the EEG generated by the generators, where the blue line in each graph is the original EEG signal, the red line is the signal generated by the depression generator, and the yellow line is the signal generated by the control generator. The four graphs in the first row are the EEG signals from the control, and it can be noticed that the output is not that good. In contrast, the four graphs in the second row are EEG signals from depressed patients, and it can be found that the depression generator fits the EEG data of depressed patients better and far outperforms the results generated by the control generator.
    From this we can determine that the depression generator learns the hidden relation between the depression and its brain activity, but the control generator does not work as well as expected.
}
    \label{4}
    \end{figure*}
        First we train a pair of learnable parameters \(\omega_{1} \quad \omega_{2}\), and sum the two feature vectors weighted by
        \(S_{hidden} = \omega_{1} * S_{cA}^{'} + \omega_{2} * S_{cD}^{'}\)
        After that, we train two Decoders, \(L_{cA}, L_{cD}\).
        Their role is to transform the intermediate hidden vectors into \(cA,cD\) features of the target signal, respectively.

        For each network, they consist of 5 nodes, while the structure of the nodes is shown in Fig.
        For the input \(S_{hidden}\) of each node, a Normalization is first performed with the following formula.
        \[
            N(S) = \frac{S-E[S]}{\sqrt{Var[S] + \epsilon}} \tag{7}   
        \]
        After that, it goes through a layer of residual connectivity with a layer of LINEAR and passes the result of the computation backward as an input to the nodes in the next layer.
        The overall formula is shown as follows.
        \[
            S_{hidden} = Linear(Linear(N(S_{hidden})) + N(S_{hidden})) \tag{8} 
        \]

        The outputs of these two networks are \(G_{cA} \in \mathbf{R}^{(1, 1255)}\),\(G_{cD} \in \mathbf{R}^{(1, 1255)}\), they represent, respectively, the wavelet coefficients of the target electrode data generated by the network \(cA^{'},cD^{'}\)
        The network is then trained by calculating the \(MSE \space Loss\) between the generated wavelet coefficients \(G_{cA},G_{cD}\) and the wavelet parameters \(O_{cA}, O_{cD}\) of the original signal
        \[
            Loss = MSE(G_{cA}, O_{cA}) + MSE(G_{cD}, O_{cD}) \tag{9}
        \]

\subsubsection[3.4.3]{Depression Discriminate}
\label{3.4.3}
For each electrode, we had two segments of generated EEG data, and by comparing which generator-generated EEG was more consistent with the original signal, we were able to determine the category of brain activity corresponding to this electrode. For each segment of EEG, we can make a depression judgement based on the number of electrodes\(n\) that match the depressive activity.
The criteria for the judgment, which we obtained in the validation set, are taken as \(n_{0}\) by sinking the distribution in the validation set, and the EEG data of \(n 
< n_{0}\) is judged as normal, while the data of \(n > n_{0}\) is judged as depressed.
        And \(n_{0}\) was chosen based on the criterion of maximizing the accuracy of the validation set.
        \begin{figure}[t]
            \centering
            \subfloat[]{
                \includegraphics[scale=0.2]{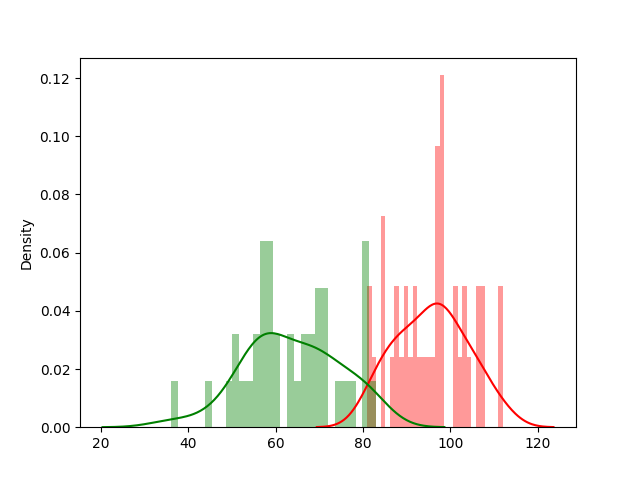}
            }
            \subfloat[]{
                \includegraphics[scale=0.2]{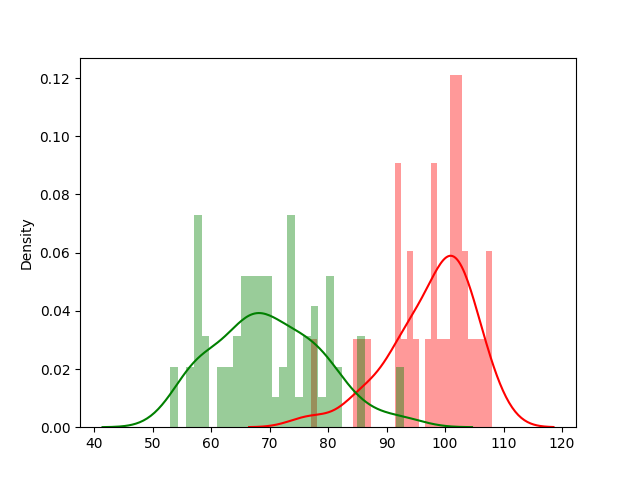}
            }
            \caption{(Fig a) is the distribution on MODMA val dataset, and (Fig b) is on test dataset. By observing the distribution in the validation set, we chose the classified line \(n_{0}=81\) and obtained the accuracy on the test set based on the criteria in the validation set.}
        \label{5}
        \end{figure}

\section{Experiments}

\subsection[4.1]{datasets and settings}
            \subsubsection[4.1.1]{dataset introduce}
            In this experiment, we used two publicly available datasets.MODMA\cite{MODMA} was composed of 
            53 participants including a total of 24 outpatients (13 males and 11 females; 16-56-year-old) diagnosed with depression, as well as 29 healthy controls (20 males and 9 females; 18-55-year-old). 
            Each subject had 8 segments of resting-state EEG data which was 10 s in length, and the instrument sampling rate was 250 Hz(\(D \in \mathbf{R}^{(8, 128, 2500)}\))
            
            The second dataset comes from\cite{HUSM}, which is the largest dataset in the world.
            This dataset has
            34 MDD outpatients (17 males and 17 females, mean age = 40.3
            ±12.9) and 30 age-matched healthy controls (21 males and 9 females, mean age = 38.3 ±15.9).
            mean age = 38.3 ± 15.6).
            Each subject had resting-state EEG data in two states, eye-open and eye-closed, and we selected EEG data from eye-closed, with a total of 29 depressed subjects and 25 controls.
            When selecting the EEG information, considering the different lengths of the EEG signals, we uniformly selected the EEG data of 20s~160s, and divided each segment of the EEG data into 14 segments of data with 10s intervals.
            The sampling frequency is 256Hz and has a total of 19 active electrodes,\(D \in \mathbf{R}{(8, 19, 2560)}\)
            \subsubsection[4.1.2]{Training data segmentation}
            When training our deep learning network, for the MODMA dataset, 
            We selected the first 15 depressed patients and control group members (with 120 segments of EEG data each) as training data, 15-20 depressed patients and control group members (with 40 segments of EEG data each) as validation dataset, and the remaining data were selected as components of the test set.
    
        For the HUSM dataset, 
            We similarly selected 1-15 depressed patients and control group members (with 210 segments of EEG data, respectively) as the training dataset, 
            the 15-20 depressed patients and control group members (with 70 segments of EEG data each) were selected as the validation dataset, and the remaining data were selected as components of the test set.
            And the remaining data were selected as components of the test set.
\subsection[4.2]{Reconstructed EEG presentation and analysis}
        (The output is shown in fig\ref{4})
        Based on the construction of the network, we expected that the signal generated by the depression generator can well fit the EEG signals of the depressed patients, the And the EEG signals comming from the control generator can well fit the EEG signals of the normal control.
        However, the experimental results are a bit different from what we expected.
        According to the final results of the experiment, we found that, on most electrodes, the signal generated by the depression generator could fit the EEG signal of the depressed patients very well. However, the output of the control generator seldom matches the control original data, and its output is more like noise compared to the performance of the depression generator.
    
        This result is acceptable. Because our starting point in designing the network was to learn the connection between EEG and brain activities.
        Obviously, doing so must presuppose a certain commonality between similar types of data. For both datasets, the subjects in the control group were randomly selected, and we can't promise that every healthy people have exactly common rule of brain activity\cite{EEG_and_Cognitive_Neuroscience}. Therefore, the results generated by the control generator are not as good as those generated by the depression generator is understandably.
        
        Also, since the depression generator comparatively well learnt characteristics of depression, on both datasets, our specificity performance is better than the sensitivity performance, which is also consistent with what we observed.
        \begin{figure}
        \centering
        \subfloat{
            \includegraphics[scale=0.2]{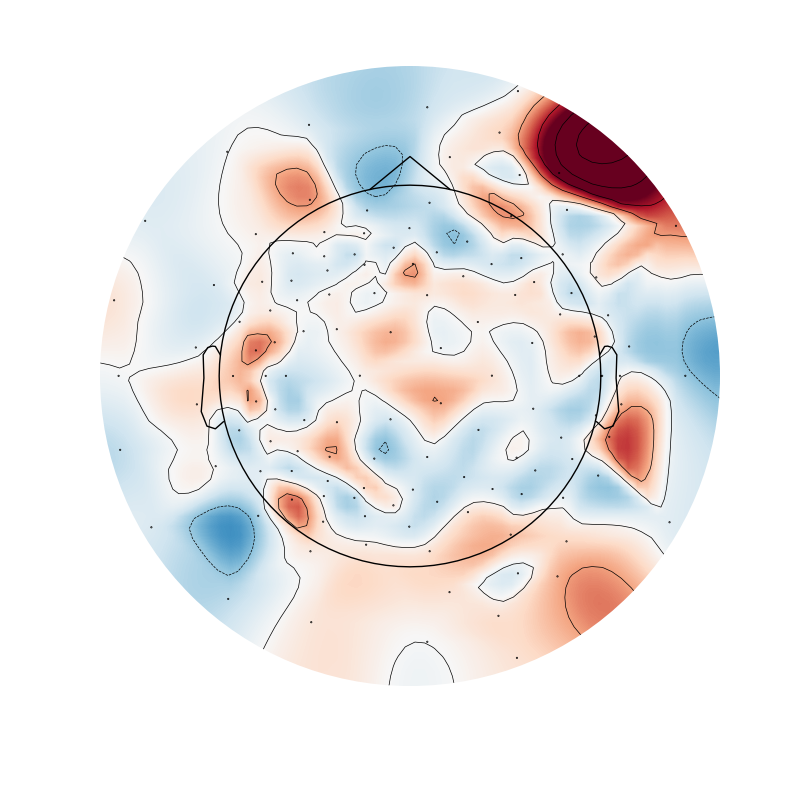}
        }
        \subfloat{
            \includegraphics[scale=0.2]{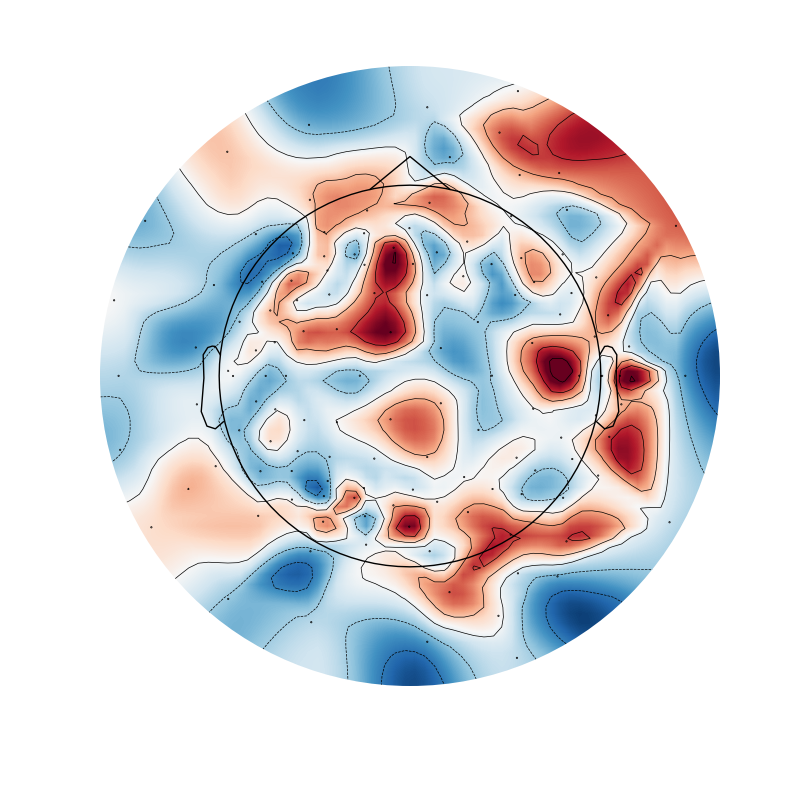}
        }
        \\
        \subfloat{
            \includegraphics[scale=0.2]{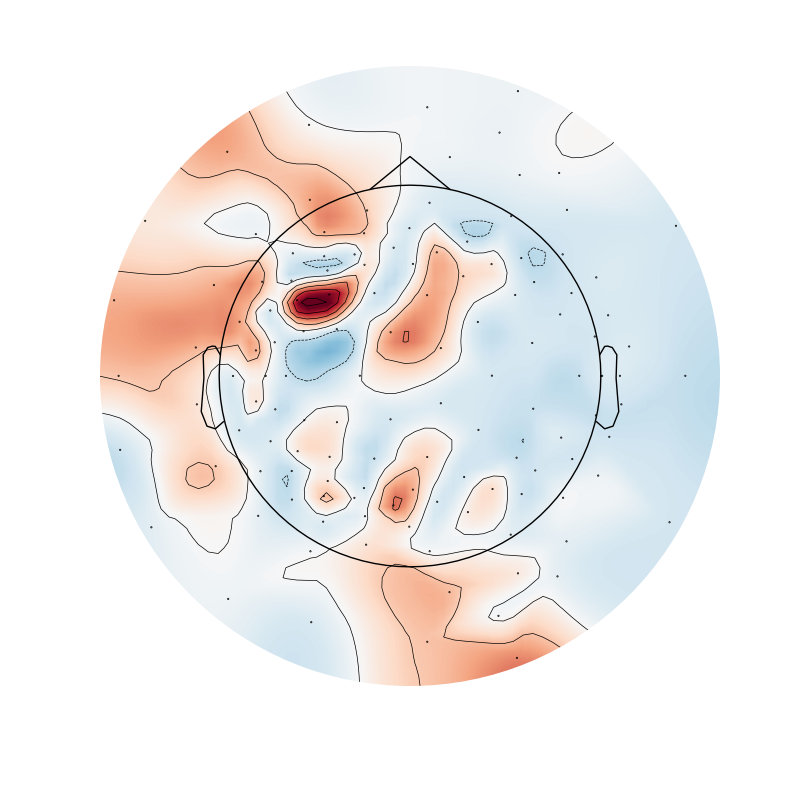}
        }
        \subfloat{
            \includegraphics[scale=0.2]{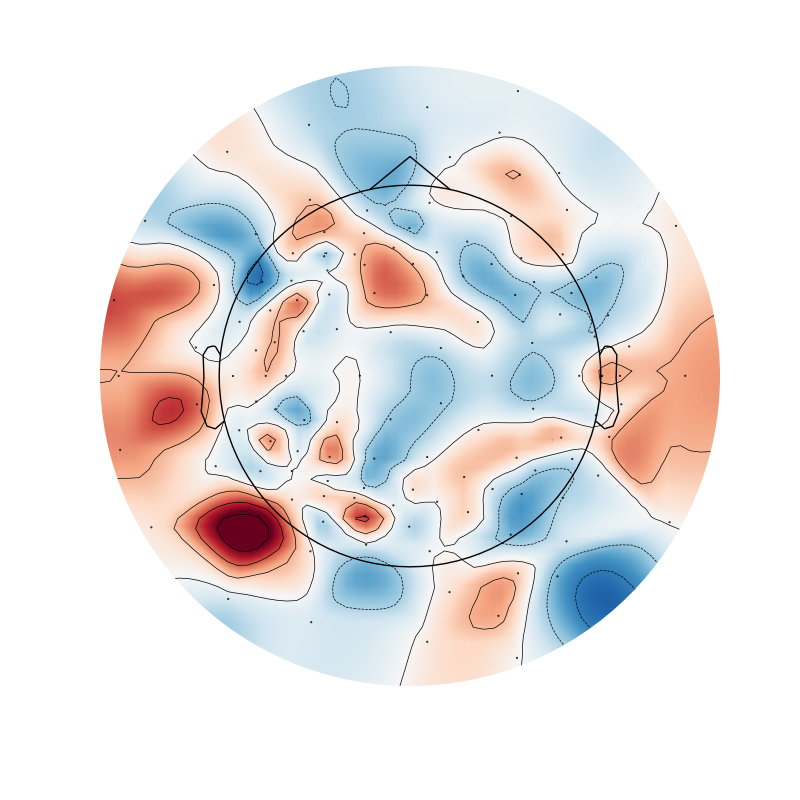}
        }
        \\
        \subfloat{
            \includegraphics[scale=0.2]{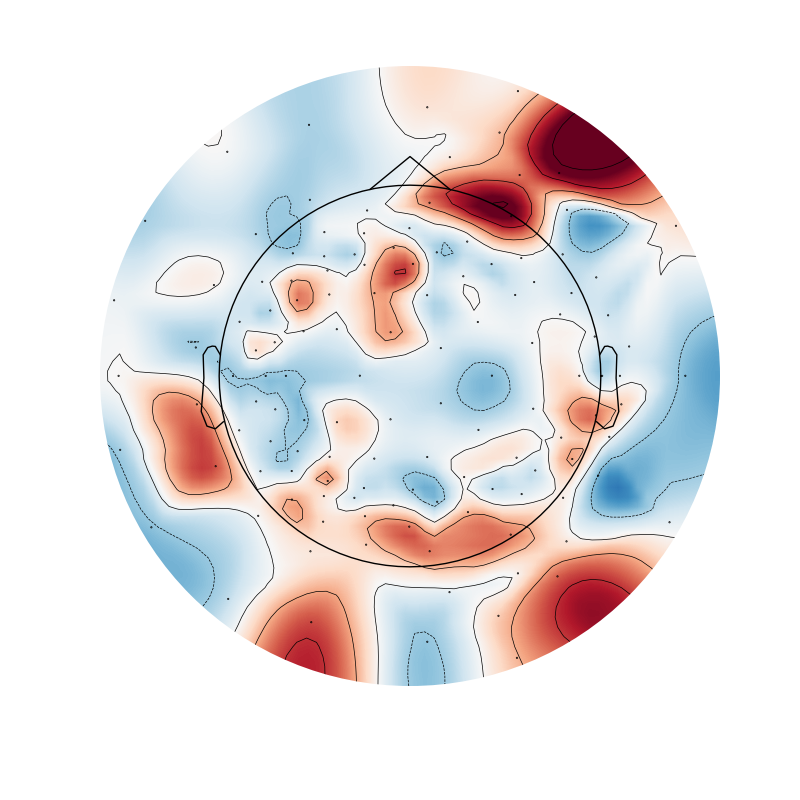}
        }
        \subfloat{
            \includegraphics[scale=0.2]{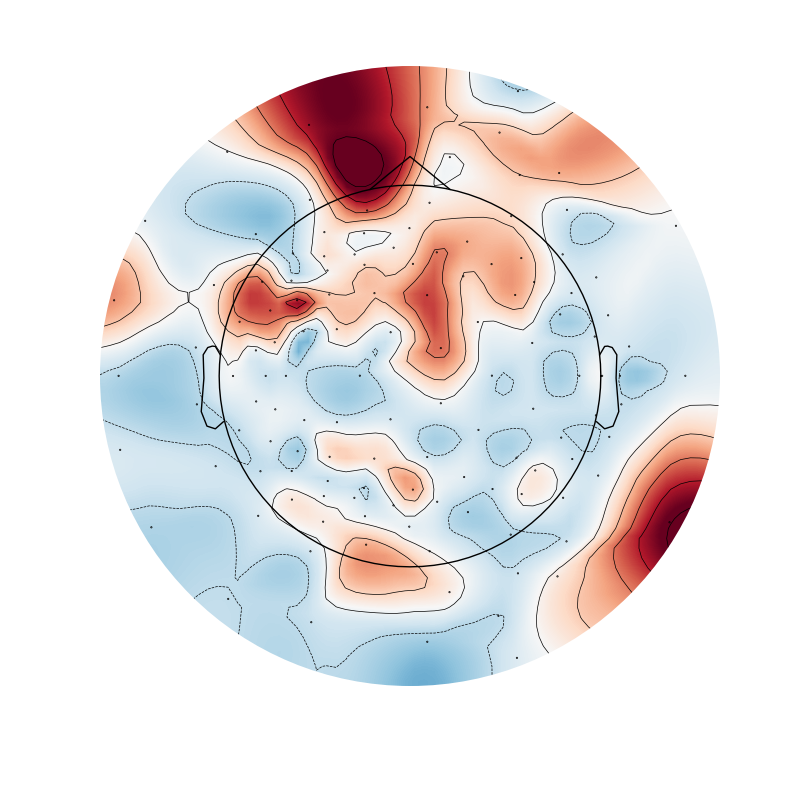}
        }
        \caption{This is a display of the final output EEG thermograms, the two pictures in the first row are the EEG topographies output when the depressive prediction is correct. The second row shows the EEG topography when the control is correctly predicted. The first picture in the third row is the EEG topography output when a depressed is misclassified as a normal, and the second is topography when a normal control is misclassified as a depression.}
        \label{7}
        \end{figure}
\subsection[4.3]{Overall Results}

As we described earlier in\ref{3.4.3}, We obtain the judgement criteria in the validation set(fig\ref{5}). On MODMA dataset, we choose \(n_{0}=81\) as judgement criteria, and
\(n_{0}=9\) on HUSM. Our model achieves highest accuracy(92.30\%) on MODMA and has the highest specificity\(96.88\%\). But on HUSM, it doesn't perform as well in MODMA, because HUSM's sigal only have 19 channels, the reconstruct of brain activity from EEG is not as accurate as in MODMA dataset.
However, the accuracy is based on several pieces of 10s-length signal from one subject. If we use multiple segments of EEG data from each subject to determine the type of subject, we achieve 100\% accuracy, on both datasets.[Show in Fig\ref{a1}]
\begin{table*}
\centering
    \begin{tabular}{p{1cm}|p{1cm}p{1cm}p{1cm}p{1cm}p{2cm}|p{1cm}p{1cm}p{1cm}p{1cm}p{2cm}}
        \toprule
        & \multicolumn{5}{c}{MODMA dataset} & \multicolumn{5}{c}{HADS dataset} \\
        \cmidrule{2-11}
        patient & class & Segment num & MDD & HC & Acc & class & Segmnent num & MDD & HC & Acc\\
        \midrule
        1 & MDD & 8 & 8 & 0 & 1.000 & MDD & 14 & 11 & 3 & 0.786\\
        2 & MDD & 8 & 8 & 0 & 1.000 & MDD & 14 & 11 & 3 & 0.786\\
        3 & MDD & 8 & 7 & 1 & 0.875 & MDD & 14 & 13 & 1 & 0.929\\
        4 & MDD & 8 & 8 & 0 & 1.000 & MDD & 14 & 14 & 0 & 1.000\\
        5 & HC & 8 & 8 & 0 & 1.000 & MDD & 14 & 13 & 1 & 0.929\\
        6 & HC & 8 & 8 & 0 & 1.000 & MDD & 14 & 12 & 2 & 0.857\\
        7 & HC & 8 & 8 & 0 & 1.000 & MDD & 14 & 13 & 1 & 0.929\\
        8 & HC & 8 & 8 & 0 & 1.000 & MDD & 14 & 13 & 1 & 0.929\\
        9 & HC & 8 & 7 & 1 & 0.875 & MDD & 14 & 13 & 1 & 0.929\\
        10 & HC & 8 & 5 & 3 & 0.625 & HC & 14 & 12 & 2 & 0.857\\
        11 & HC & 8 & 7 & 1 & 0.875 & HC & 14 & 11 & 2 & 0.857\\
        12 & HC & 8 & 7 & 1 & 0.875 & HC & 14 & 12 & 2 & 0.857\\
        13 & HC & 8 & 7 & 1 & 0.875 & HC & 14 & 11 & 2 & 0.857\\
        14 & \textbackslash & \textbackslash & \textbackslash & \textbackslash & \textbackslash & HC & 14 & 11 & 2 & 0.857\\
        \midrule
                Result & \multicolumn{5}{c}{MODMA dataset accuracy=100\%} & \multicolumn{5}{c}{HADS dataset accuracy=100\%}\\

        \bottomrule
        
    \end{tabular}
    \caption{For multiple segments of EEG data for each subject, our judgements are shown above. It can be seen that for each subject, our model is able to accurately judge whether he is depressed or not.}\label{a1}
    \end{table*}
\subsection[4.4]{Explainable Information}
\begin{table}
        \centering
            \begin{tabular}{p{2cm}|p{1.5cm}p{1.5cm}p{1.5cm}}
                \toprule & \multicolumn{3}{c}{MODMA dataset} \\
                \cmidrule{2-4}
                    Models& Sensitivity & Specificity & Average\\
                \midrule
                \cite{Shen2021An} & \textbackslash & \textbackslash & 81.60\%\\
                \cite{EEGNet} & 91.45\% & 75.60\% & 84.34\%\\
                \cite{23EP} & 88.0\% & 82.5\% & 85.3\%\\
                \cite{Shenjian23} & 87.50\% & 89.17\% & 88.50\%\\
                \textbf{Ours} & \textbf{96.88}\% & \textbf{90.28}\% & \textbf{92.30}\%\\
                \bottomrule
            \end{tabular}
            \caption{Experiments on MODMA dataset}\label{table1}
            \begin{tabular}{p{2cm}|p{1.5cm}p{1.5cm}p{1.5cm}}
                \toprule & \multicolumn{3}{c}{HUSM dataset} \\
                \cmidrule{2-4}
                    Models& Sensitivity & Specificity & Average\\
                \midrule
                \cite{HUSM} & 95\% & 80\% & 87.5\% \\
                \cite{EEGNet} & 88.74\% & 87.97\% & 88.36\%\\
                Ours & 89.68\% & 81.43\% & 86.73\%\\
                \bottomrule
            \end{tabular} 
            \caption{Experiments on HUSM dataset}\label{table2}
            
        \end{table}
        We hold it in mind that it is difficult to rely on a single algorithm to fully
        solve the depression recognition problem. 
        Therefore, we design networks that not only require them to be physiologically logical, but also to output a certain amount of explainable information to reduce misjudge. 
        For different electrodes, the generated data does not match the original data to the same degree.
        Some areas fit better, while others fit poorly. 
        Thus, we can plot the EEG heatmap based on the difference
        between the generated and the original.
        The more similar the generated signal in a particular region is to the original signal,
        the more similar the features of the brain activity in this region are to the features learnt by the generator.
        Areas with better conformity have high values, by looking at whether the areas with highs value echo some of the current research consensus, for example whether the distribution is more focused on 
        the right-side hemisphere, frontal and parietal-occipital cortex\cite{EEG_biomarker2}, it is possible to tell if this result is a misjudgement(see fig\ref{7}).
\subsection{Number of channel selected}

In our previous experiments, we selected the 10 most similar electrodes, meanwhile, we conducted a comparison test in MODMA dataset that 5,10,15,20 electrodes were selected for the experiments, and the following is the table of our experimental results.
We found that at 5-10 electrodes, as the number of electrodes rises, the accuracy of judgment rises, 
and while at 10-20 electrodes, the accuracy is decreasing.[see table\ref{table3}]

\begin{table}

\begin{tabular}{p{2cm}|p{1cm}p{1cm}p{1cm}p{1cm}}
                \toprule
                channel number & 5 & 10 & 15 & 20 \\
                \midrule
                    Sensitivity &100.0\% & 96.9\%& 100.0\% & 12.5\% \\
                    Specificity & 79.2\% & 90.3\%& 54.2\% & 19.4\%\\
                    Accuracy    & 85.6\% & \textbf{92.3}\%& 68.3\% & 17.3\%\\
                \bottomrule         
            \end{tabular}
            \caption{Table of number of similar electrodes and final accuracy}\label{table3}
\end{table}
\section[5]{Conclusion}
In this article, we present an Generative Detection network(GDN) for detecting depression, aiming to make a depression diagnosis through the characteristics of brain activity.
We trained two generators, one is trained totally on depression dataset and another is on the normal control's, and we expect that each generator could learn the connection between EEG signals and brain activity in a particular category, so that the generator could generate a target electrode signal by several signals 'similar' to it. 
The construct of the model consistents with three physiological laws, which makes our model's output more convicing and reliable.
As a result, our modelachieves high accuracy in experiments to determine the category of EEG segments that are 10s long(92.3\% on MODMA and 86.73\% on HUSM), and we successfully judged the category of subjects in each test.
Besides, our network could put out explainable information which could help us determine whether to trust the results of the network, and we found that the explainable information echoing the findings of the current study, reinforcing the reliability of our network.

\bibliographystyle{named}
\bibliography{ijcai24}
\end{document}